# A Workflow-Forecast Approach To The Task Scheduling Problem In Distributed Computing Systems

Andrey Gritsenko, postgraduate student

Department of Informational Technologies, Institute of Mathematics and Natural Sciences
North-Caucasus Federal University
Stavropol, Russia

*Abstract* — **The aim of this paper is to provide a description of deep-learning-based scheduling approach for academic-purpose high-performance computing systems. Academic-purpose distributed computing systems' (DCS) share reaches 17.4% amongst TOP500 supercomputer sites (15.6% in performance scale) that make them a valuable object of research. The core of this approach is to predict the future workflow of the system depending on the previously submitted tasks using deep learning algorithm. Information on predicted tasks is used by the resource management system (RMS) to perform efficient schedule.**

## I. INTRODUCTION

The analysis of an academic-purpose Zewura cluster's workload described in [1] reveals a great amount of periodically submitted tasks with similar resource requirements and duration (Fig. 1). The subsequent inquiry of Zewura workload carried out by the author made it possible to ascertain that periodic patterns of tasks were submitted by a certain group of users –students. This unveiling allowed to make an assumption that such reiterative nature of workload is inherent for every academic-purpose cluster system. Additional examination of the workload of the CIS's biggest supercomputer system Lomonosov has disclosed same recurring patterns of jobs and thus confirmed that assumption.

It becomes obvious that seriate task submissions allow to use predictive methods to forecast the future workflow of the cluster system. First of all to perform any of the predictive algorithms a list of submitted tasks should be presented as a time series. In articles [2, 3] the transformation process of the workload into the time series is described. Time series contains the following information about jobs – submission time and resource requirements: number of requested nodes and processor time.

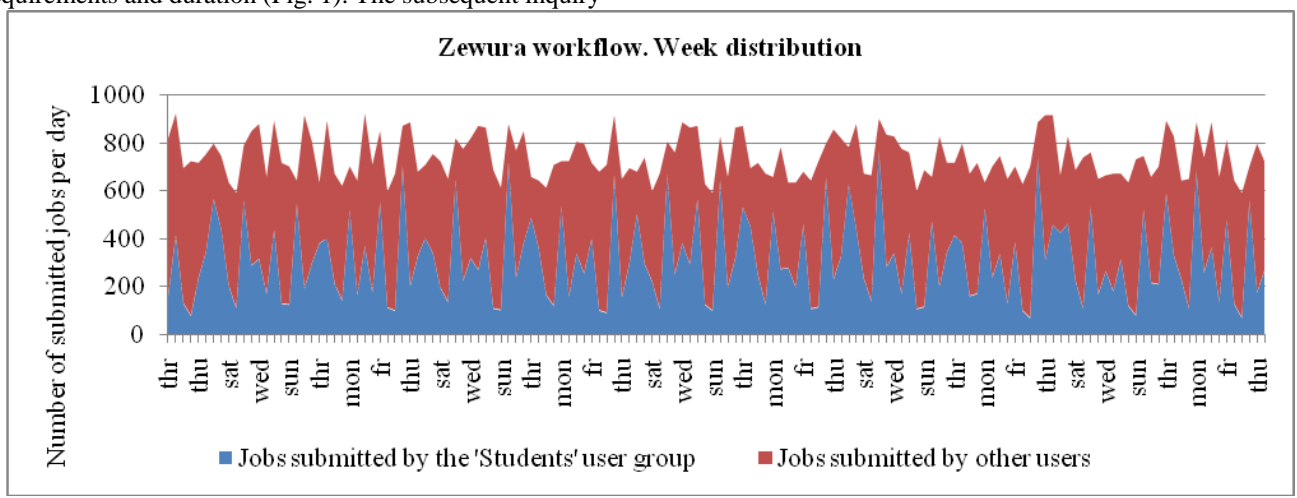

Figure 1.   Recurring patterns of the tasks submitted by the users of the Student user group observed in the Zewura workflow.

To obtain more reasons for using forecasting methods to predict the future workload for academic-purpose high-performance systems the Hurst exponent [4] for both Zewura and Lomonosov workflow time series was computed. The Hurst exponent is used as a measure of long-term memory of time series, in other words it determines the rate at which autocorrelation function of time series decreases as the lag between pairs of values increases. The presence of long-term memory that corresponds to the value of Hurst exponent $H \in (0.5;1)$ in turn makes the prognosis of future tasks highly reliable. The Hurst coefficient computed for Zewura workflow amounted to 0.714731 and for Lomonosov cluster system's set of jobs – 0.69814 thereby endorsing relevance of prediction algorithms employment.

## II. USE OF DIFFERENT WORKFLOW PREDICTION MODELS

Before the deep-learning algorithm was applied to solve the prediction problem some other forecasting methods were also tested in assumption they could provide a relevant forecast.

Research described in [5] is devoted to the appliance of ARIMA [6] and GMDH [7] methods as well as SSA [8] forecasting algorithm in order to gain a proper forecast for the Zewura workflow. Fig. 2 depicts the comparison of the dispersion of predictions obtained via using different models

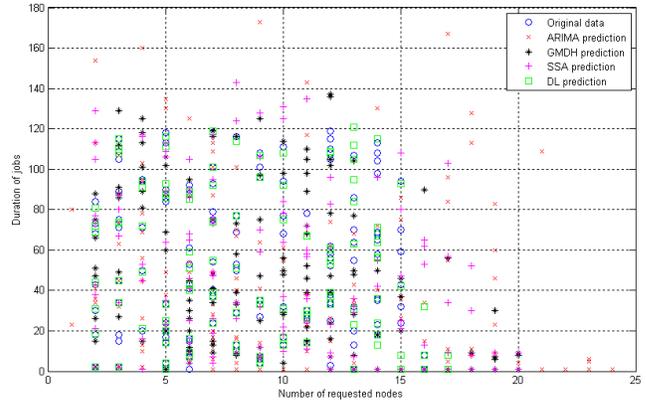

Figure 2. Dispersion of the results of different forecasting methods in processor-time space. Each value on the figure represents a submitted task that has certain resources requirements and computed for a certain time.

As one can see from the Fig. 2 forecasted values of every method deviate both in time and nodes dimensions except for the Deep Learning (DL) approach: for this algorithm most of the predicted values slightly deviate mainly in one dimension – job duration.

Fig. 3 additionally shows a great inaccuracy of the results when using algorithms mentioned above.

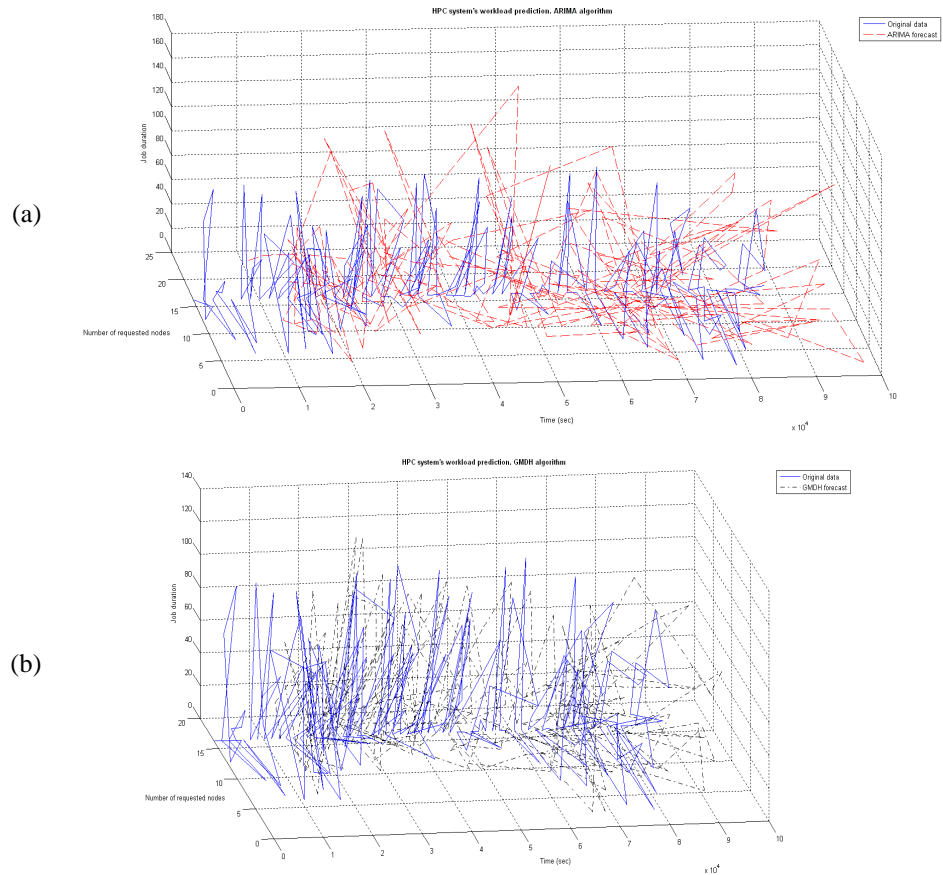

(a)

(b)

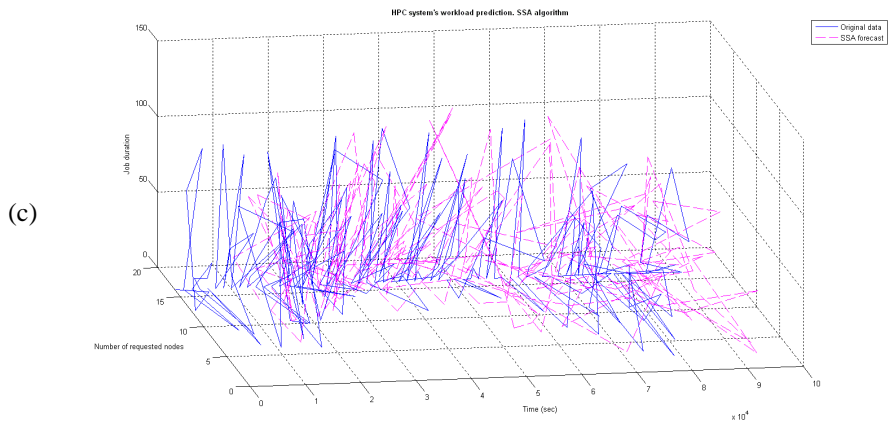

Figure 3. Prediction inaccuracy of different forecasting algorithms: (a) ARIMA method, (b) GMDH method, (c) SSA metnod.

## III. DEEP-LEARNING-BASED FOREACSTING APPROACH

To solve the contradiction consisting in simultaneous presence of both long-term memory and low efficiency of examined forecasting methods it was proposed to decompose time series to reveal essential seriated components. The deep learning approach was chosen to implement for concurrent periodic components extraction and their subsequent prediction. The core of the deep learning approach is to build a multi-layer structure of features where each additional layer is formed on the basis of the previous and the initial data are an input for the lowest layer [9]. The process of the deep-learning-based decomposition is described in more detail in [10] and is performed as follows.

Reasoning from the deep learning approach concept initial features that would be used to decompose time series should be firstly derived. According to the figured problem the periodicity of tasks seems to be the most relevant and thus was chosen as the initial features where each new feature layer corresponds to the patterns of jobs with longer (less) periodicity. For instance, the first layer consists merely of all of the submitted tasks and the second layer represents patterns of similar jobs that recur during the day. In this way the highest layer would represent patterns of repeating groups of jobs, for example, a group of jobs that recur every day at 2 p.m. during a month, where the group itself recur every half-year. Fig. 4 visually illustrates the process of feature layers detection. In addition it should be noticed that jobs' revelation process is mostly reminds a simple search of the tasks with similar requirements such as number of requested computational resources and processor time submitted by a certain user of group of users.

An irrefutable advantage of the deep-learning approach in comparison with previously mentioned forecasting algorithms (ARIMA, GMDH, SSA) is the lack of necessity to perform the prediction of the further workflow in the truest sense of the word: revealed seriate components could be simply prolonged in the future. The recurring jobs revealed by means of the utilization of the described deep-learning approach for the workflow prediction are represented on the Fig. 5 as blue circles together with the predicted values illustrated as red crosses.

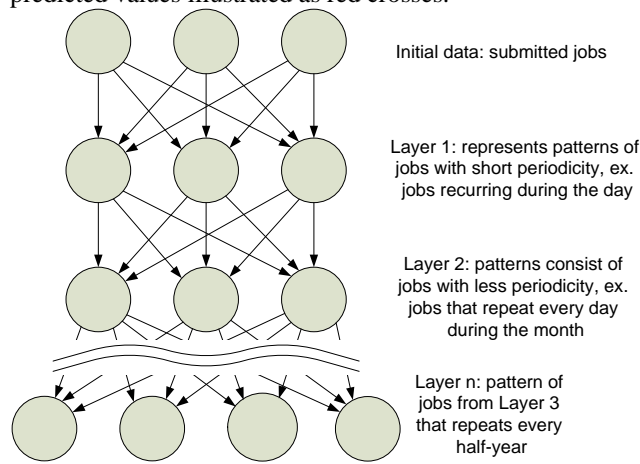

Figure 4. Deep learning layer detection for workflow prediction.

On the other hand this deep-learning approach where prediction as such does not apply but instead revealed patterns of recurring tasks are prolonged in the future sets an additional problem that could be defined as follows: how long should we continue revealed patterns of submitted similar tasks?

To manage with this problem a specific decision-making system was designed [11]. The main idea on which the functioning of the system is based is to use the position of the predictive value in the pattern comparative to the average length of patterns consisted of jobs with similar parameters.

When one get a new forecasted task for a certain pattern of recurring jobs he should determine a group of patterns with similar periodicity and resembling job requirements. Then an expectation value and a standard deviation for this group of patterns thereby constructing normal distribution should be computed.

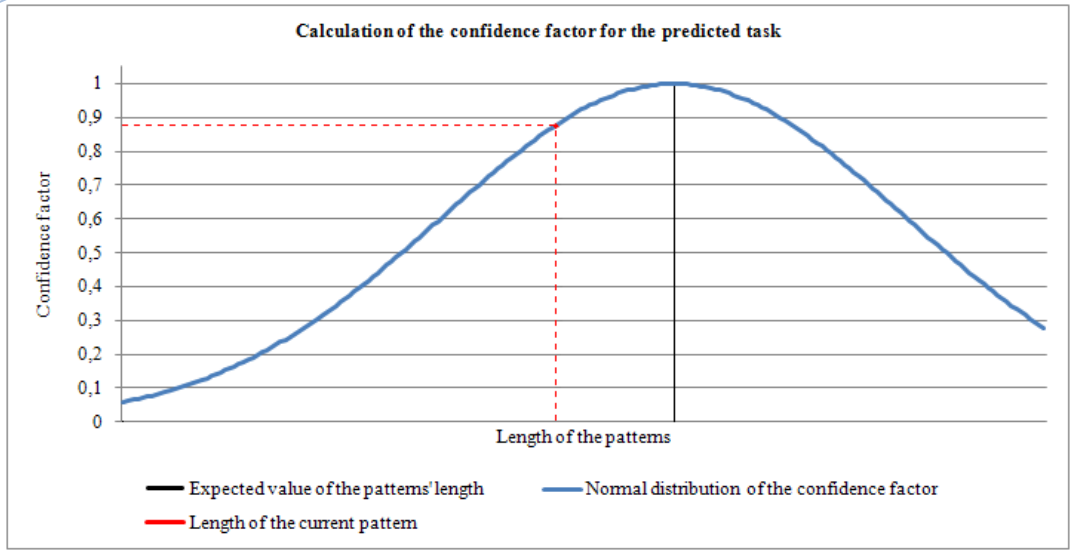

Figure 5. Normalized confidence factor for a predicted task of a certain pattern of jobs. Black line determines the expected value of the lengths of the patterns with similar periodicity and job parameters. Red dot depicts the value of the confidence factor calculated for a pattern of a certain length

Once a normal distribution was obtained and normalized a confidence factor could be computed for every forecasted task in the following way: first of all the length of the pattern, i.e. number of tasks in the pattern, is calculated taking into account the predicted job as well. Depending on the length of the pattern a corresponding confidence factor could be obtained (Fig. 5). The aim of this factor is to show the likelihood of the predicted job to come true, in other words that the forecasted task would be submitted.

The next step the system decides what to do with the predicted task. Depending on the value of the confidence factor there are several opportunities:

- in case of small values of the confidence factor the decision-making system ignores information about the predicted task;
- for medium values of the confidence factor the system makes a decision to reserve resources required for task computation, though this reservation could be freed in order to compute jobs with higher priority;
- reservations made for tasks with high values of the confidence factor could not be freed or altered in any other way.

The advantage of the deep-learning approach is that it could modify the borders for 'small', 'medium' and 'high' ranges for confidence factor values during its work depending on the provided data. For instance, if a predicted task with a low value of the confidence factor came true then deep-learning algorithm would decrease both the upper bound of the 'small' range and the lower bound of the 'medium' range of jobs, and on the contrary, if a predicted task with a high value of the confidence factor has not come true that would increase the borders of 'high' and 'medium' ranges.

## IV. MULTI-CRITERIA EVALUATION PROCESS

Before comparing the efficiency of different scheduling methods and algorithms one should at first determine objective functions. There is a plenty of different criteria to assess the efficiency but the following three were chosen as the most representative:

### A. Makespan

Makespan is an essential objective function that is defined as the completion time of the last task in the workflow. In comparison to the flow time this criterion is turned out to be unbiased as it does not depend on the execution order of tasks.

### B. Computational resource utilization

Instead of estimating the efficiency of a schedule calculating merely the percentage of used CPUs, one should compute the resource utilization using the following expression that permits not to consider situations when there are not enough submitted jobs to use all the available resources:

$$Resource\ utilization = \frac{CPU_{active}}{\min\left(CPU_{available}; CPU_{requested}\right)}$$

Computational resource utilization makes it possible to assess the efficiency of a schedule in respect to the economic aspect of the scheduling problem.

### C. Slowdown

Slowdown objective function is a dimensionless quantity that is calculated as follows:

$$Slowdown = \sum_{Tasks} \frac{FinishTime - SubmitTime}{FinishTime - StartTime}$$

Slowdown also involves both wait time (difference between start time and submission time) and response time

(difference between finish time and submission time) and its advantage in comparison to them is that slowdown also takes into account processing time of each task thus decreasing the influence of the small tasks being in the queue for a long time.

Objective functions that anywise evaluate fairness of tasks distribution over the resources or tardiness have not been considered because due to the TOP500 statistics most of the academic-purpose computational systems are homogeneous, in other words constructed of identical nodes, and in academic-purpose system's workloads (including Zewura and Lomonosov) tasks are submitted without due date till which they should be completed.

The multi-criteria evaluation process could be divided in two steps where the first step is to assign weights to every objective function. A binary preference matrix, or binary comparison matrix, the most convenient way to compare criteria in multiple objective decision problems, was used to prioritize chosen objective functions. Table I contains results of binary comparisons of criteria: 1 if row objective is more preferable then column objective and 0 vice versa, if both criteria equivalently preferable then the respective value in the matrix is 0.5; the ranks of criteria are presented in the forth column and calculated as the sum of values in the corresponding row.

The decision on such values of the binary comparison matrix could be explained by means of the following reasons.

- From the point of view of the users the slowdown objective is much more important than the resource usage as no one wants its task to be in the queue for a long time.
- The aim of the distributed computing system is to complete all the tasks in the queue as possible in a short period of time therefore the makespan is as significant criterion as the slowdown, besides frequently these objective functions correlate in a way where decrease of the slowdown results in the decrease of the makespan.
- On the other hand, taking into consideration high cost of the distributed systems' equipment the value of the resource utilization objective increases drastically making its influence on the overall schedule efficiency rating more significant.

TABLE I.        BINARY COMPARISON MATRIX

| Criteria | Makespan | Slowdown | Resource usage | $\Sigma$ |
|---|---|---|---|---|
| **Makespan** | – | 0.5 | 0 | 0.5 |
| **Slowdown** | 0.5 | – | 1 | 1.5 |
| **Resource usage** | 1 | 0 | – | 1 |

The next step of the multi-criteria evaluation process consists in the comparison itself of various scheduling algorithms. The comparison process proposed in [12] performs an assessment of algorithms based on the denoted objective functions:

*1) Max and min values of objectives:* For every objective the biggest and the smallest values are determined on the set of specified algorithms.

*2) Relative estimations for objectives:* For each algorithm a relative estimation is calculated as a ratio of the difference between algorithm's objective value and the min value of this objective to the difference between max and min values of the same objective. This estimation shows the proximity of the evaluated algorithm to the best algorithm on a set of objectives.

*3) Global estimations:* The next step is to compose a square matrix with the number of columns and rows equals to the number of compared algorithms. Each element of the matrix represents the ratio of the overall superiority of the row algorithm on the column algorithm to the overall superiority of the column algorithm on the row algorithm.

*4) Final estimations:* To define which of the compared scheduling algorithms performs the most efficient schedule on a set of specified objective functions one should calculate a main eigen vector for the defined prefence matrix. An algorithm that relate to the biggest element of the main eigen vector is the desired algorithm.

## V. COMPARISON OF DIFFERENT SCHEDULING ALGORITHMS

To test the performance of the proposed deep-learning approach for the task scheduling problem a grid simulation system Alea 3.1 [13] that allows to perform a various number of scheduling algorithms of both queue-based and schedule-based approaches was used. The main distinguish between these approaches is that algorithms of the schedule-based approach require information about all tasks in the workflow at the moment of scheduling, in other words perform a static scheduling, while queue-based algorithms perform dynamic scheduling when a schedule is constructed every time a new task is submitted. This feature of the schedule-based approach makes it impossible to use its algorithms in real distributed computing systems though allows using them in simulators, especially taking into account that average values of the objective functions for schedule-based methods is often higher than those for queue-based methods.

To perform as possible exhaustive and reliable comparative study the following algorithms were specified along with the proposed deep-learning algorithm (DL):

- queue-based: First-Come-First-Served (FCFS), Smallest-Job-First (SmJF), Conservative Backfilling (Cons BF), Aggressive /EASY Backfilling, Last-Come-First-Served (LCFS), Shortest-Job-First (ShJF), First-Fit, Earliest-Deadline-First (EDF) algorithms used in PBS-Pro [14] resource management system;

- schedule-based: Earliest-Suitable-Gap (ESG), Best-Gap, Tabu-Search.

Mentioned above Zewura and Lomonosov workflows were used as the input data for the experiments. The results of the execution of different algorithms scheduling Zewura and Lomonosov workloads are presented in the Tables II, III, IV and V respectively.

TABLE II.     Absolute Values Of The Objective Functions For Different Scheduling Algorithms. Zewura Workflow

| Criteria | Scheduling algorithms | | | | | | | | | | | | | |
| | FCFS | Smallest JF | Cons BF | EASY BF | LCFS | SJF | First Fit | PBS-Pro | BSG | ESG | DL | Best Gap | Tabu Search | EDF |
|---|---|---|---|---|---|---|---|---|---|---|---|---|---|---|
| Makespan | 16630898 | 16410022 | 14977165 | 15224472 | 15754907 | 16159979 | 16109705 | 15214316 | 15365688 | 15074443 | 15223893 | 15372390 | 15698837 | 16630898 |
| System usage | 77.016 | 85.044 | 90.989 | 87.931 | 82.744 | 86.088 | 79.82 | 88.3 | 86.653 | 92.642 | 92.284 | 92.749 | 83.738 | 77.016 |
| Slowdown | 10916.33 | 956.9934 | 493.1645 | 1111.561 | 1904.009 | 1184.289 | 6960.194 | 465.7136 | 576.358 | 261.7794 | 257.016 | 238.4517 | 1933.402 | 10916.33 |

TABLE III.     Relative Global Estimations For Most Efficient Algorithms. Zewura Workflow

| | Scheduling algorithms | | | | Main eigen vector |
| | Cons BF | DL | BestGap | TabuSearch | |
|---|---|---|---|---|---|
| Cons BF | 1 | 0.053464727 | 0.146786008 | 0.2 | 0.038 |
| DL | 18.7039206 | 1 | 13.99131226 | 1.902052263 | 0.9474 |
| BestGap | 6.812638428 | 0.071472924 | 1 | 0.502942619 | 0.1391 |
| Tabu Search | 5 | 0.525747909 | 1.988298389 | 1 | 0.2856 |

TABLE IV.     Absolute Values Of The Objective Functions For Different Scheduling Algorithms. Lomonosov Workflow

| Criteria | Scheduling algorithms | | | | | | | | | | | | | |
| | FCFS | Smallest JF | Cons BF | EASY BF | LCFS | SJF | First Fit | PBS-Pro | BSG | ESG | DL | Best Gap | Tabu Search | EDF |
|---|---|---|---|---|---|---|---|---|---|---|---|---|---|---|
| Makespan | 107100566 | 107108258 | 107098023 | 107097817 | 107099830 | 107101707 | 107100548 | 107099792 | 107100516 | 107098164 | 107098060 | 107098124 | 107100516 | 107100566 |
| System usage | 98.45 | 99.027 | 98.753 | 98.547 | 98.69 | 98.974 | 98.785 | 99.131 | 98.997 | 98.89 | 98.965 | 99.051 | 98.623 | 98.45 |
| Slowdown | 612.998 | 69.33852 | 212.874 | 151.0045 | 208.2882 | 254.5014 | 596.2115 | 186.2868 | 289.0596 | 76.40185 | 88.26473 | 90.6587 | 309.628 | 612.998 |

TABLE V.     Relative Global Estimations For Most Efficient Algorithms. Lomonosov Workflow

| | Scheduling algorithms | | | | | | Main eigen vector |
| | SmallestJF | EASY BF | PBS-Pro | DL | BestGap | TabuSearch | |
|---|---|---|---|---|---|---|---|
| SmallestJF | 1 | 3.738761168 | 2.570680305 | 0.672726853 | 0.714458873 | 0.519489967 | 0.1813 |
| EASY BF | 0.267468275 | 1 | 0.547116287 | 0.010761058 | 0.007653458 | 0.008980744 | 0.0143 |
| PBS-Pro | 0.389002086 | 1.827764998 | 1 | 0.277450973 | 0.212094044 | 0.104856179 | 0.0585 |
| DL | 1.486487414 | 92.92766715 | 3.604240378 | 1 | 1.140552745 | 0.658719961 | 0.4249 |
| BestGap | 1.399660691 | 130.659892 | 4.714889595 | 0.876767869 | 1 | 0.229324149 | 0.4442 |
| TabuSearch | 1.924964993 | 111.3493506 | 9.53687245 | 1.51809579 | 4.360639751 | 1 | 0.7653 |

## VI.     Conclusion

As it can be observed from the tables above the deep-learning-based algorithm described in this article offers a great performance compatible to the performance of the schedule-based methods and superior to the ones of the queue-based methods.

In spite of this fact the area of application of the deep-learning-based scheduling algorithm is constrained a few since there are several restrictions and requirements placed upon distributed computing systems in order to implement the proposed method properly.

The main restriction consists in the requirement of the academic purpose of distributed computing systems that is imposed in order to guarantee the presence of the patterns of the recurring jobs. The future work on the analysis of workloads of non-academic cluster systems should be done with the purpose of revealing recurring tasks that in their turn would allow to apply the deep-learning-based scheduling algorithm.

The list of minor restrictions placed on distributed computing systems involves homogeneous architecture of these systems and non-preemptive job scheduling. In modern cluster systems with heterogeneous architecture there are certain queue associated with certain type of computational resources. Consequently each heterogeneous system could be considered as a set of several homogeneous systems that at first glance allow to evade the restriction. On the other hand the strong heterogeneity is essential to grid systems as well as preemptive job scheduling that is crucial to these systems. Taking into account all of the above it could be stated that redemption of both homogeneous and non-preemptive restrictions would expand the area of application on grid systems.

Thus there is some future research to be done to make the proposed deep-learning-based algorithm widely applicable for the task scheduling problem for distributed computing systems with different principles of operation and architectures.

## ACKNOWLEDGMENT

The Zewura workload log was graciously provided by the Czech National Grid Infrastructure MetaCentrum [15].

The Lomonosov workload was gained gathering information on submitted tasks via monitoring the official site that shows current state of the cluster system Lomonosov. [16].

Alea 3.1 software is the result of the research intent No. 0021622419 (Ministry of Education, Youth and Sports of the Czech Republic) and the grant No. 201/07/0205 (Grant Agency of the Czech Republic). The owner of the result is Masaryk University, a public high school, ID: 00216224.